\documentstyle[12pt]{article}

\def\thefootnote{\fnsymbol{footnote}}
\def\tF{{\tilde F}}
\def\tM{{\tilde M}}
\def\notp{\not{\hspace{-.03in}p}}

\def\notcD{\not{\hspace{-.05in}{\cal D}}}
\def\notF{\not{\hspace{-.05in}F}}
\def\notD{\not{\hspace{-.05in}D}}

\def\bea{\begin{eqnarray}}
\def\eea{\end{eqnarray}}
\def\lll{{\ln\Lambda^2\over 32\pi^2}}
\def\ibar{\bar{\imath}}
\def\jbar{\bar{\jmath}}

\def\ba{\bar{\alpha}}

\def\bmu{\bar{\mu}}
\def\[{\left [}
\def\]{\right ]}
\def\({\left (}
\def\){\right )}
\def\lbr{\left\{}
\def\rbr{\right\}}
\def\pp{\partial}
\def\M{\bar{M}}

\def\z{\bar{z}}

\def\R{{\cal{R}}}
\def\S{{\cal{S}}}

\def\STr{{\rm STr}}
\def\Tr{{\rm Tr}}

\def\G{{\cal G}}

\def\tcF{{\tilde\F}}

\def\H{{\cal H}}
\def\L{{\cal L}}
\def\D{{\cal D}}
\def\bl{\bar{\lambda}}
\def\hf{\hat{\phi}}
\def\hz{\hat{z}}
\def\hM{\hat{M}}
\def\hF{\hat{F}}
\def\hA{\hat{A}}

\def\bF{\bar{F}}

\def\W{\overline{W}}
\def\hV{\hat{V}}
\def\n{\bar{n}}
\def\m{\bar{m}}
\def\s{\bar{s}}
\def\cM{{\cal{M}}}
\def\bl{\bar{\lambda}}

\def\bc{\bar{\chi}}

\def\tcF{{\tilde{\cal F}}}
\def\A{\bar{A}}
\def\bc{\bar{\chi}}
\def\bps{\bar{\psi}}
\def\bl{\bar{\lambda}}

\def\theequation{\ksection.\arabic{equation}}

\catcode`\@=11

\def\thesubsection{\Alph{subsection}.}
\def\thesubsubsection{\arabic{subsubsection}.}
\begin{document}

\begin{titlepage}
\begin{center}

\hfill    LBL-34457 \\
\hfill    UCB-PTH-93/22 \\
\hfill    August, 1993

\vskip .2in

{\large \bf SUPERGRAVITY COUPLED TO CHIRAL MATTER AT ONE LOOP}\footnote{This
work was supported in part by
the Director, Office of Energy Research, Office of High Energy and Nuclear
Physics, Division of High Energy Physics of the U.S. Department of Energy under
Contract DE-AC03-76SF00098 and in part by the National Science Foundation under
grants PHY--90--21139.}
\vskip .15in

Mary K. Gaillard {\em and} Vidyut Jain\footnote{Address after
September 1, 1993:
 Inst. for Theoretical Physics, SUNY at Stony Brook, NY 11794}\\[.03in]

{\em  Department of Physics, University of California and\\
      Physics Division, Lawrence Berkeley Laboratory, 1 Cyclotron Road,\\
      Berkeley, California 94720}\\[.02in]

\end{center}

\vskip .05in
\begin{abstract}
%insert abstract here

We extend earlier calculations of the one-loop contributions to the
effective bose Lagrangian in supergravity coupled to chiral matter.
We evaluate all logarithmically divergent contributions for arbitrary
background scalar fields and
space-time metric. We show that, with a judicious choice of gauge
fixing and of the definition of the action expansion, much of the result can
be absorbed into a redefinition of the metric and a renormalization of the
K\"ahler potential.  Most of the remaining terms depend on the curvature of
the K\"ahler metric. Further simplification occurs in
models obtained from superstrings in which the K\"ahler Riemann tensor is
covariantly constant.

\end{abstract}
%%%%%%%%
\newpage
%%%%%%%%%%%%%%%%%%%%%%%%%%%%%%%%%%%%%%%%%%%%%%%%%%%%%%%%%%%%%%%
\renewcommand{\thepage}{\roman{page}}
\setcounter{page}{2}
\mbox{ }
\vskip 1in
\begin{center}
{\bf Disclaimer}
\end{center}
\vskip .2in
\begin{scriptsize}
\begin{quotation}
This document was prepared as an account of work sponsored by the United
States Government. Neither the United States Government nor any agency
thereof, nor The Regents of the University of California, nor any
of their employees, makes any warranty, express or implied, or assumes
any legal liability or responsibility for the accuracy, completeness, or
usefulness of any information, apparatus, product, or process disclosed,
or represents that its use would not infringe privately owned rights.
Reference herein to any specific commercial products process, or service by its
trade name, trademark, manufacturer, or otherwise, does not necessarily
constitute or imply its endorsement, recommendation,or favoring by the
United States Government or any agency thereof, or The Regents of the
University of California. The views and opinions of authors expressed herein
do not  necessarily state or reflect those of the United States Government
or any agency thereof of The Regents of the University of California and
shall not be used for advertising or product endorsement purposes.

\end{quotation}
\end{scriptsize}
\vskip 2in
\begin{center}
\begin{small}
{\it Lawrence Berkeley Laboratory is an equal opportunity employer.}
\end{small}
\end{center}
\end{titlepage}

\newpage
\renewcommand{\thepage}{\arabic{page}}
\setcounter{page}{1}
\def\thefootnote{\arabic{footnote}}
\setcounter{footnote}{0}

\section{Introduction}\indent

    Considerable progress has recently been made in understanding Yang-Mills
couplings at the quantum level~\cite{anomalies},~\cite{tom}
in effective supergravity theories
obtained from superstrings.  Specifically, it is understood how to cancel the
modular anomaly that arises at the quantum level of the effective field theory.
{}From the field theory point of view, the modular anomaly is equivalent to the
standard chiral and conformal anomalies of Yang-Mills theories. In particular,
the conformal anomaly enters through the dependence of the effective cut-off
on the moduli fields~\cite{tom},~\cite{bg}.   In a general field theory the
conformal
anomaly entails all operators that have logarithmically divergent coefficients
at the quantum level.  Understanding the structure of the divergences in the
full effective supergravity theory is a necessary step in determining what
counterterms are needed to fully restore modular invariance.
The determination of
these loop corrections may also provide a guide to the construction of an
effective theory for a composite chiral multiplet that is a bound state of
strongly coupled Yang-Mills superfields, which in turn could shed light on
gaugino condensation as a mechanism for supersymmetry breaking.

    In References~\cite{josh},~\cite{noncan} we identified the divergent
one-loop contributions to the effective bose Lagrangian, with a flat space-time
background metric, in a general $N=1$ supergravity theory, with specialization
to the no-scale form suggested by superstrings. Here we present the full
results for a general supergravity theory coupled to chiral matter with an
arbitrary background space-time metric and arbitrary background scalar fields.
Partial results for a curved-space time
metric have been given in~\cite{sigma},~\cite{sred}, and particularly
in~\cite{lahanas}, where it was shown how to recast the
Einstein term in canonical form by a redefinition of the background metric.
However, the results are gauge dependent~\cite{gauge}, and therefore not very
meaningful
unless one can isolate those terms that actually contribute to the S-matrix.
This is the purpose of the present paper.  We choose a gauge fixing
prescription which, together with a redefinition of the expansion of the
action,
enhances supersymmetry cancellations between
boson and fermion loop contributions.
With these choices, all operators of dimension six or less, and most of those
of
dimension eight, that do not depend on the K\"ahler curvature can be either
absorbed by field redefinitions or interpreted as renormalizing the K\"ahler
potential.  By an operator of dimension $d$ we mean a K\"ahler invariant
operator whose term of lowest dimension is $d$, where scalar fields are
assigned
the canonical dimension of unity.  In many effective theories from
superstrings,
such as the untwisted sector in many orbifold compactifications, the K\"ahler
Riemann tensor is covariantly constant; in this case the results simplify
further.

In order to complete the program of determining one-loop supergravity, the
Yang-Mills sector must be included.  We will present the full results in a
subsequent paper~\cite{future}, where we will also consider the parity odd
operators that arise from integration over fermionic degrees of freedom.
As mentioned above, the effective cut-off of effective theories
derived from superstrings is field dependent; moreover the field dependence is
different for loop corrections arising from different sectors of the
theory~\cite{tom}.  Here we use a single cut-off and neglect its derivatives.
The latter does not represent a
loss of generality, since terms involving derivatives of the cut-off have a
different dependence on the moduli and must be considered together with terms
that are one-loop finite. Our results, some of which are collected
an appendix, will be presented in such a way that the contributions from
different sectors can be isolated and the correct cut-offs included.

In Section 2 we discuss gauge fixing and describe the prescription used here.
The results of our calculation are presented in a succinct form in Section 3;
further simplifications arising in models from string theory are pointed out
in Section 4.  In Appendix A we define our conventions and give that part of
the tree-level Lagrangian that is needed to perform our calculations.  In
Appendix B we list the operators that appear in the quantum action as defined
by our gauge fixing and expansion prescriptions, as well as the traces of
products of these operators that determine the divergent terms in the
effective one loop action. In a final appendix we list corrections to and
misprints in~\cite{josh},~\cite{noncan}.

\section{Gauge Fixing and the Expansion of the Action}
\setcounter{equation}{0}\indent

The S-matrix is independent of gauge fixing and also of shifts in the
propagators that are proportional to $\L_A = \pp\L/\pp\phi^A$ where $\phi^A$
is any field.  However, certain choices can lead to an effective Lagrangian
that
better displays the symmetries of the theory.  For example, we expand the
action
$S$ in terms of normal scalar coordinates~\cite{normal},~\cite{mkg} $\hz^I$:
\begin{equation}
S = S(z) + D_IS\big|_z\hz^I + {1\over 2}D_ID_JS\big|_z\hz^I\hz^J + \cdots,
\end{equation}  where $D_I$ is the field redefinition covariant derivative
defined in appendix A, and interpret the determinant of the second term in
(2.1)
as the one-loop effective action for a scalar theory.  The result differs
from that of a standard Taylor expansion by
terms of the form $F^{JL}(z)\Gamma(z)^I_{JK}(D_IS)_z$, where $\Gamma^I_{JK}$ is
the connection associated with the covariant derivative $D_I$, and $F^{JL}$ is
an arbitrary matrix-valued function of the background scalar fields.  Such
terms
vanish when the classical equations of motion for the background fields $z$
are satisfied. The expansion (2.1) yields an effective action that is
manifestly field redefinition invariant.  It therefore preserves nonlinear
symmetries among the scalar fields, up to quantum anomalies.

Supersymmetry is also a nonlinear
symmetry in supergravity theories, even when auxiliary fields are used.
We have no formal argument by which we can determine the gauge fixing and
expansion prescription so as to yield an effective action that is manifestly
supersymmetric.\footnote{Since we set background fermions to zero,  our
effective action cannot be manifestly supersymmetric.  However supersymmetry
constrains~\cite{cremmer},~\cite{kahler} the bosonic part of the action; by
``manifest supersymmetry'' we are referring to these constraints.}
Instead, we adopt a pragmatic approach, and
use prescriptions that give the most boson-fermion cancellations, and/or
simplify the calculation.  We find that with our prescription the operators of
dimension six or less can be interpreted as renormalizations of the tree
Lagrangian, except for those that depend on the scalar curvature tensor.
Additional operators of dimension eight can be isolated into terms of the form
$F^{JL}\Gamma^I_{JK}D_IS|_z$, which do not contribute to the S-matrix.
It turns out that the gauge fixing prescription that satisfies these properties
yields an effective quantum Lagrangian that is of a particularly simple form:
all the propagators are
the same as those of standard scalar or spin-${1\over 2}$ fermions.  It is
possible that this feature contributes to the enhanced cancellations.
We first discuss the case of flat SUSY Yang-Mills theory, where a similar
gauge fixing dependence arises~\cite{barb}, and where a ``supersymmetric
gauge'' can be found.

\subsection{Supersymmetric Yang-Mills theory}\indent

In background field calculations of the effective one-loop action, the Landau
gauge fixing condition $\D^\mu\hA_\mu = 0$ has frequently been
used~\cite{josh},~\cite{noncan},~\cite{sigma}.  In the absence of a
superpotential, the dimension four operators of the resulting supergravity
Lagrangian for the gauge nonsinglet scalars
can be interpreted in terms of two renormalizations. The first is a
renormalization of the matrix-valued function
$x^a_b(z,\z) = $ Re$f(z)^a_b$ that normalizes the Yang-Mills kinetic term
$-{1\over 4}x^a_bF^b_{\mu\nu}F_{a\mu\nu}$.  The second is a renormalization
of the K\"ahler potential $K(z,\z)$, where $z=(\z)^{\dag}$ is a
complex scalar field.  Here (and throughout) we
consider the case $x^a_b = \delta^a_bx$ at tree level, for which
the results are:
\begin{equation}
\delta K = \lll\[-{2\over x}K_{\m j}(T_a\z)^{\m}(T^az)^j\] +
{\rm higher\; dimension\; terms},
\end{equation}
where $T^a$ represents the gauge group on the scalar field $z^n =
(\z^{\n})^{\dag}$, and
\begin{equation}
\delta x_a^b = \lll\[2D_i(T_az)^jD_j(T^bz)^i - 6C_{G}^{(a)}\delta^b_a\] +
{\rm higher\; dimension\; terms},
\end{equation}
where $C_{G}^{(a)}$ is the Casimir of the adjoint representation and the
field redefinition covariant scalar derivative $D_i$ is defined in Appendix A.
The fact that (2.3) is not the real part of a holomorphic function has been
discussed elsewhere in the literature (see, {\it e.g.},~\cite{anomalies}). In
the flat SUSY limit $x\to$ constant, $K_{i\m}\to \delta_{im}$, and the
renormalizations reduce to constants that depend on the Casimirs of the matter
representations $R$:
$$ \delta K_{i\m} \to -{\ln\Lambda^2\over 16\pi^2 x}\sum_a(T_a)^2_{i\m} =
-\delta_{im}{\ln\Lambda^2\over 16\pi^2 x}\sum_aC_2^a(R_i), $$
$$ \delta x_a^b = \delta^a_b{\ln\Lambda^2\over 16\pi^2 x}\Tr(T_a)^2 =
\delta^a_b{\ln\Lambda^2\over 16\pi^2 x}\sum_RC^a_R. $$

When a superpotential is included, the results obtained in the Landau gauge
can no longer be
interpreted in terms of these renormalizations. This is similar to the result
found in~\cite{barb}.  However, if we use a smeared gauge fixing prescription
defined by
\begin{equation}
\L\to\L - {x\over 2}C_aC^a, \;\;\;\; C^a = \D^\mu\hA^a_\mu +
{i\over x}\[(T^a\z)^{\m}\hz^i - (T^a\z)^i\hz^{\m}\]K_{i\m},
\end{equation}
the results can once again be interpreted as above, with, instead of (2.2),
\begin{equation}
\delta K = \lll\(-{4\over x}K_{\m j}(T_a\z)^{\m}(T^az)^j +
e^{-K}A_{ij}\A^{ij}\) + {\rm higher\; dimension\; terms},
\end{equation}
where $A_{ij}$ is defined in Appendix A; in the flat SUSY limit it reduces to
the second derivative of the superpotential $W$:
$$ e^{-K}A_{ij}\A^{ij}\to e^KW_{ij}\W^{ij}. $$
Note that the gauge-dependent term in (2.5) differs by a factor of two from
that
in (2.2).  The result (2.5) agrees with the chiral matter wave function
renormalization found in~\cite{barb} and in a recent string loop
calculation~\cite{ant}.

Unlike the Landau gauge, the smeared gauge fixing (2.4)
gives a quantum Lagrangian of the simple form (3.1) below.  The
field-dependent masses have the correct poles for unitarity when evaluated at
the ground state configuration for the background fields, {\it i.e.,}
$\D_\mu z = A_\mu = \pp_i V = 0,$ where $V$ is the scalar potential.
  We will use gauge fixing prescriptions
for supergravity that share this feature.
In addition, the transformation laws for supergravity
are nonlinear even when auxiliary fields are used.\footnote{Once the auxiliary
fields have been eliminated, the transformation laws for fermions in flat
supersymmetry are also nonlinear. However it is easy to show that
eliminating these fields before or after functional integration gives the same
one-loop effective action.}  This suggests that it may be necessary to
redefine~\cite{gauge} the expansion in a manner analogous to (2.1), in order
to obtain a manifestly SUSY result.

\subsection{Gauge-fixing the gravity supermultiplet}\indent

We set background fermions to zero, and use unhatted symbols for
quantum fermion fields ($\psi,\chi,\lambda$).

The commonly used gauge fixing\footnote{The gauge fixing of supergravity using
superfields is considered in~\cite{super}, where it is necessary to introduce
``ghosts of ghosts'' because the Faddeev-Popov action has itself a gauge
invariance, as well as so-called ``hidden'' ghosts because the gauge smearing
parameters are constrained.  The component action gauge fixing we describe here
has no such proliferation of ghosts.} for the
graviton~\cite{tini},~\cite{vid},~\cite{josh},~\cite{lahanas}, when
generalized to include the YM sector, is defined by
\begin{equation}
\L \to\L + {1\over 2}C_\mu C^\mu, $$
$$ C_\mu = {1\over\sqrt{2}}\(\nabla^\nu h_{\mu\nu}
- {1\over 2}\nabla_\mu h^\nu_\nu - 2\D_\mu z^IZ_{IJ}\hz^J + x
F^a_{\mu\nu}\hA_a^\nu \),
\end{equation}
where $Z_{IJ}(z,\z)$ is the scalar metric, $\hz,\;\hA$ are the quantum scalar
and gauge fields, and
the symmetric tensor $h_{\mu\nu}$ is the quantum part of the
gravitational field.  Like the smeared Yang-Mills gauge fixing (2.4), this
leads
to a Lagrangian of the form (3.1).

For the gravitino, two types of gauge fixing have been used: the Landau
gauge~\cite{ino},~\cite{josh} $\gamma\cdot\psi = 0$, which is implemented with
the aid of an auxiliary field, and the smeared gauge fixing~\cite{lahanas}
$\L\to\L - \bF\cM F,\; F = \gamma\cdot\psi, \; \cM = {1\over 4}\(i\notD
+ 2M_\psi\),$ which requires Nielsen-Kallosh ghosts. Neither of these has
the feature that the quantum Lagrangian reduces to the simple form (3.1).
In addition, while the Landau gauge propagators have the correct poles for
constant background fields, the smeared gauge fixing propagators do not.
Here we adopt an unsmeared gauge which satisfies both requirements.

In a supergravity theory in which the Yang-Mills normalization function
satisfies Re$f_{ab} = \delta_{ab}x$, the part of the Lagrangian that depends
on the gravitino $\psi_\mu$ is~\cite{cremmer},~\cite{kahler}
\bea
\L_\psi &=& {1\over 4}\bps_\mu\gamma^\nu(i\notD + M)\gamma^\mu\psi_\nu -
{1\over 4}\bps_\mu\gamma^\mu(i\notD + M)\gamma^\nu\psi_\nu
+\bigg[{x\over 8}\bps_\mu\sigma^{\nu\rho}\gamma^\mu\lambda_aF^a_{\nu\rho}
\nonumber \\
 & & - \bps_\mu\notcD \z^{\m}K_{i\m}\gamma^\mu L\chi^i +
{1\over 4}\bps_\mu\gamma^\mu\gamma_5\lambda^a\D_a
-i\bps_\mu\gamma^\mu L\chi^im_i + {\rm h.c.}\bigg] \nonumber \\
 & & + {\rm four-fermion \; terms}.\eea
where
\begin{equation}
\M = (M)^{\dag} = e^{K/2}\(WR + \W L\), \;\;\;\; R,L =
{1\over 2}\(1\pm\gamma_5\)$$
$$ m_n = (\m_{\n})^{\dag} = e^{-K/2}D_i(e^KW),\;\;\;\; \D_a = K_i(T_a z)^i.
\end{equation}
We take the Landau gauge condition $G=0$, where
\bea
G & =& -\gamma^\nu(i\notD - \M)\psi_\nu
- {x\over 2}\sigma^{\nu\rho}\lambda_aF^a_{\nu\rho} \nonumber \\ & &
- 2(\notcD z^iK_{i\m}R\chi^{\m} + \notcD\z^{\m}K_{i\m}L\chi^i) + 2im_I\chi^I
- \gamma_5\D_a\lambda^a
\eea
which we implement by inserting a $\delta$-function in the functional integral
over $\hf$.  Writing
$$\delta[G-G(\hf)] = \int d\alpha\;{\rm exp}\(i\alpha[G-G(\hf)]\), $$
and defining
$$ \psi' = \psi +\gamma\alpha, \;\;\;\;\bps' = \bps + \ba\gamma, $$
We obtain
\bea
\L &=& -{1\over 2}\bps'^\mu(i\notD - \M)\psi'_\mu +
{1\over 2}\ba\gamma^\mu(i\notD - \M)\gamma_\mu\alpha +
{\rm matter \; terms} \nonumber \\
&=& -{1\over 2}\bps'^\mu(i\notD - \M)\psi'_\mu -\ba(i\notD + 2M)\alpha
+ \ba\({x\over 2}\sigma^{\nu\rho}\lambda_aF^a_{\nu\rho}
+ 2im_I\chi^I - \gamma_5\D_a\lambda^a\) \nonumber \\
& & -ix\bps'_\mu\notF^\mu_a\lambda^a
- 2\bps'_\mu(\D^\mu\z^{\m}K_{i\m}L\chi^i + \D^\mu z^iK_{i\m}R\chi^{\m}).
\eea
Note that $\psi$ is C-even: $\psi=C\bps^T$, then $\psi'=C\bps'^T$
requires $\alpha = -C\ba^T$, {\it i.e.} $\alpha$ is C-odd; note also
that $\alpha$ has
negative metric.\footnote{In the notation of (3.1), $Z_{\alpha\alpha}= -2$;
including the contribution proportional to Det$Z_{\alpha\alpha}$ we get a
quartically divergent term proportional to
$\ln 2$ which cancels a similar contribution from the graviton
ghost~\cite{josh}.} All the terms remaining in the Lagrangian (2.10) are of the
form of either a mass or a connection; that is, (2.10) is of the form (3.1).

To obtain the ghostino determinant we use the supersymmetry
transformations~\cite{cremmer}
\begin{equation}
i\delta\psi_\mu = (iD_\mu - {1\over 2}\gamma_\mu M)\epsilon, \;\;\;\;
i\delta\chi^i = {1\over 2}(\notcD z^iR - i\m^iL)\epsilon, $$
$$ i\delta\chi^{\m} = \[{1\over 2}(\notcD\z^{\m}L - im^{\m}R)\]\epsilon,
\;\;\;\; \m^i = K^{i\m}\m_{\m}, \;\;\;\; m^{\m} = K^{i\m}m_i, $$
$$ i\delta\lambda^a = \[-{i\over 4}\gamma^\mu\gamma^\nu F^a_{\mu\nu}
- {1\over 2x}\D^a\]\epsilon,
\end{equation}
to obtain
\bea
{\pp\delta G\over\pp\epsilon} &=& D^\mu D_\mu -
{1\over 2}\gamma^\mu\gamma^\nu[D_\mu,D_\nu] - i[\notD,M]- 2M\M + \m^im_i + \D
\nonumber \\
& & +2i\m_{\m}\notcD\z^{\m}L + 2im_i\notcD z^iR
+{x\over 2}\sigma_{\sigma\rho}F_a^{\sigma\rho}
[{1\over 4}\sigma^{\mu\nu}F^a_{\mu\nu} + {1\over x}\gamma_5\D^a] \nonumber\\
& & - \D_\mu z^iK_{i\m}\D^\mu\z^{\m}
 + {1\over 2}\gamma_5[\gamma^\mu,\gamma^\nu]\D_\mu\z^{\m}K_{i\m}\D_\nu z^i.
\eea
For constant background fields the ghostino propagator becomes
\begin{equation}
 D^\mu D_\mu - 2M\M + \m^im_i + \D = D^\mu D_\mu + M\M + V,
\end{equation}
where $V$ is the potential. When we evaluate this at a ground state with a
flat background metric, the vacuum energy necessarily vanishes: $V=0$, so
the (4-fold) ghostino pole is at $p^2 = - D^2 = M^2$.  If the cosmological
constant is nonzero the curvature is also, and there are additional terms in
all the masses.

Now the goldstino is unmixed with the gravitino, but instead mixes with
$\alpha$. The normalized (left handed) goldstino field $\chi_L$ is
\begin{equation}
\chi_L = \(m_i\chi^i_L - {i\over 2}\D_a\lambda^a_L\)
\bigg/\sqrt{{1\over 2}(m_i\m^i + \D)},
\end{equation}
and its mass is
\begin{equation}
m_\chi = e^{-K}\(e^{-K}A_{ij}\A^i\A^j + 4\D\A - {1\over 2x}\D f_i\A^i\)
\bigg/\(e^{-K}A_i\A^i + \D\),
\end{equation}
where $\lambda$ is a gaugino, $\chi^i$ is the left-handed superpartner of
$z^i$,
and
$$A_{ij} = D_iA_j = D_iD_jA, \;\;\;\; A = e^KW =\M. $$
At the ground state
\begin{equation}
V_i = 0 = \A^iV_i =
e^{-K}A_{ij}\A^i\A^j + 2\D\A - {1\over 2x}\D f_i\A^i - 2e^{-K}\A^iA_i\A.
\end{equation}
Using this gives $m_\chi = 2M$.

Here we show that unitarity is satisfied
in the case where there are no gauge couplings:  $\D_a=0$;  the argument goes
through in the same way when gauge couplings are included~\cite{future}.
The normalized~\cite{josh} $(\alpha,\chi)$ mass matrix is
\begin{equation}
\M_{1\over 2} = \pmatrix{M^i_{\m} & M^i_\alpha\cr M^\alpha_{\m} &
M^\alpha_\alpha\cr} = \pmatrix{ \bar{\mu}^i_{\m} & i\m^i\cr i\m_{\m} &
-2\M\cr}, \end{equation}
where
$$ \mu_{ij} = (\bmu_{\ibar\jbar})^{\dag} = e^{-K/2}A_{ij}, \;\;\;\;
\bar{\mu}^{ij} = K^{i\m}K^{j\n}\bar{\mu}_{\m\n},$$
is the normalized mass matrix for left-handed chiral fermions.
In the traces used to evaluate the one-loop effective action (see Section 3)
this gets multiplied by
\begin{equation}
M_{1\over 2} = \pmatrix{M^{\m}_j & M^{\m}_\alpha\cr M^\alpha_j &
M^\alpha_\alpha\cr} = \pmatrix{\mu^{\m}_j & im^{\m}\cr im_j & -2M\cr},
\end{equation}
so
\begin{equation}
\M_{1\over 2}M_{1\over 2} = \pmatrix{\bar{\mu}^{ik}\mu_{kj} -\m^im_j &
i\bar{\mu}^{ik}m_k - 2iM\m^i \cr i\m^k\mu_{kj} - 2i\M m_j & 4M\M - \m^km_k\cr}.
\end{equation}
For the Goldstino at a ground state with vanishing cosmological constant [see
(2.13), (2.16)], $\mu_{ij}\to 2M,\;m_i\to\sqrt{3}M$, so
the $\alpha$-Goldstino squared mass matrix reduces to
$$ \M_{1\over 2}M_{1\over 2} \to M\M\pmatrix{1&0\cr0&1}. $$
The ghostino determinant removes four poles at $p^2 = M^2$ and the unphysical
fields $\chi$ and $\alpha$ restore two of them,
so the singularities are correct.  Note that because $\alpha$ is C-odd
while $\chi$ is C-even, $M_{1\over 2}$, which operates on left-handed fermions
is not the hermitian conjugate of $\M_{1\over 2}$, which operates on
right-handed fermions.

\subsection{Modification of the graviton propagator}\indent

The S-matrix is unchanged if we add terms proportional to $\L_A$ to the
propagators, as in (2.1).  Consider the graviton-scalar sector.  We have:
\bea \L_i &=& -\(K_{i\m}d^\mu\D_\mu\z^{\m} + V_i\), \nonumber \\
\L_{\mu\nu} &=& {1\over 2}g_{\mu\nu}\({r\over 2}-V +
K_{i\m}\D_\rho z^i\D^\rho\z^{\m}\) \nonumber \\ & &
  \quad - {1\over 2}r_{\mu\nu}
- {1\over 2}\(K_{i\m}\D_\mu z^i\D_\nu\z^{\m} + K_{i\m}\D_\nu
z^i\D_\mu\z^{\m}\) \nonumber \\
\L_\mu^\mu &=& {r\over 2}-2V + K_{i\m}\D_\rho z^i\D^\rho\z^{\m}.
\nonumber \eea
where $g_{\mu\nu}$ is the background metric.
We can redefine the graviton propagator by:
\begin{equation}
\Delta^{-1}_{I\mu\nu} =
{1\over\sqrt{g}}\(D_I D_{\mu\nu}S - {1\over 2}g_{\mu\nu}D_IS\),
\end{equation}
and
\bea
\Delta^{-1}_{\mu\nu,\rho\sigma} &\to& \Delta^{-1}_{\mu\nu,\rho\sigma}
- P_{\mu\nu,\rho\sigma}\L^\lambda_\lambda
-{1\over 2}\[g_{\mu\nu}\L_{\rho\sigma} + g_{\rho\sigma}\L_{\mu\nu}\]
\nonumber \\
& & + {1\over 2}\[g_{\mu\rho}\L_{\nu\sigma} + g_{\nu\rho}\L_{\mu\sigma} +
g_{\mu\sigma}\L_{\nu\rho} + g_{\nu\sigma}\L_{\mu\rho}\]
\equiv - \(P\nabla^2 + X\)_{\mu\nu,\rho\sigma}, \nonumber \\ & & \quad
\eea
where the spin-2 projection operator $P$ is defined in (B.2), and
$$ \L_{\mu\nu} = g_{\mu\rho}g_{\nu\sigma}{\pp\L\over\pp g_{\rho\sigma}}.$$
The unmodified propagators have been evaluated\footnote{As a check, we have
also calculated the curvature dependent terms using the unmodified propagators;
we agree with the results of~\cite{tini}, but not with~\cite{lahanas} for these
terms.} elsewhere~\cite{tini},~\cite{vid}; using these
results in the above we get
\begin{equation}
X_{\mu\nu,\rho\sigma} = -2P_{\mu\nu,\rho\sigma} V
- {1\over 4}\[r_{\mu\rho\nu\sigma} + r_{\nu\rho\mu\sigma}\]
\end{equation}

Evaluating the determinants in (3.2) below gives an effective Lagrangian
including terms linear and quadratic in the space-time curvature:
\begin{equation}
\L_1\ni \L_r = {1\over 2}\sqrt{g}\[\epsilon_0(z,\z)r + H_{\mu\nu}\(\D_\rho z,
\D_\rho\z, F_{\rho\sigma}\)r^{\mu\nu} + \alpha r^2 + \beta
r^{\mu\nu}r_{\mu\nu}\].
\end{equation}
The Einstein term can be put in canonical form by a redefinition of the
metric~\cite{lahanas}:
\bea
g_{\mu\nu}&=& (1 - \epsilon)g^R_{\mu\nu} + \epsilon_{\mu\nu},
\nonumber \\
\epsilon & =& \epsilon_0 + \epsilon_\alpha, \;\;\;\;
\epsilon_{\mu\nu} = \H_{\mu\nu} -{1\over 2}g_{\mu\nu}\H^\lambda_\lambda ,
\nonumber\\
 \epsilon_\alpha &=& \alpha\(r + 4V\) + \beta V, \nonumber \\
\H_{\mu\nu} &=& H_{\mu\nu} - \beta g_{\mu\nu}{x\over 4}F_{\sigma\rho}
F^{\sigma\rho} - 2\alpha g_{\mu\nu}\D_\rho z^i\D^\rho\z^{\m}K_{i\m}
\nonumber \\
& & \!\!\!\!\!\! + \beta\(r_{\mu\nu} -\D_\mu z^i\D_\nu\z^{\m}K_{i\m} -
\D_\nu z^i\D_\mu\z^{\m}K_{i\m} + xF_{\mu\rho}F^{\;\;\rho}_\nu\).
\eea
This induces additional matter terms:
\bea  & &
\L(g) + \L_1(g) = \L(g_R) + \L_1 - \L_r + \sqrt{g}\bigg(
2\epsilon V - \epsilon \D_\mu z^i\D^\mu\z^{\m}K_{i\m} \nonumber \\
& & \qquad + {1\over 2}\H^\mu_\mu V - \H^{\mu\nu}\D_\mu z^i\D_\nu\z^{\m}K_{i\m}
+{x\over 2}\H^{\mu\nu}F_{\mu\rho}F^{\;\;\rho}_\nu -
{x\over 8}\H^\nu_\nu F_{\mu\rho}F^{\mu\rho}\bigg),\nonumber
\\ && \quad \eea
where the tree Lagrangian $\L(g)$ is given in Appendix A.
Note that any terms containing factors of $\L_{\mu\nu}$ that
can appear in $\L_1$ are
completely removed by this metric redefinition.

\section{The One-Loop Effective Action}
\setcounter{equation}{0}\indent

In the absence of gauge fields, the quantum action obtained by the
prescriptions defined in the preceeding section takes the form
\begin{equation}
\L_q = -{1\over 2}\Phi^T Z_\Phi\(D^2 + H_\Phi\)\Phi + {1\over 2}
\bar{\Theta}Z_\Theta\(i\notD - M_\Theta\)\Theta + \L_{gh} + \L_{Gh}.
\end{equation}
The last two terms are the ghost and ghostino terms, respectively, $\Phi =
(h_{\mu\nu},\hz^i,\hz^{\m})$ is a $2N+10$ component scalar, $\Theta =
(\psi_\mu, \chi^I = L\chi^i + R\chi^{\ibar}, \alpha)$ is an $N + 5$ component
Majorana fermion, where $N$ is the number of chiral multiplets, and the matrix
valued metrics $Z_\Phi$ and $Z_\Theta$, as well as the matrix-valued covariant
derivative $D_\mu$, are defined in Appendix A.
The one loop contribution to the effective action is
\bea
\L_1 &=& {i\over 2}\Tr\ln(D^2 + H_\Phi) -{i\over 2}\Tr\ln(-i\notD + M_\Theta)
\nonumber \\
& & + i\Tr\ln(D^2 + M^2_{Gh}) - i\Tr\ln(D^2 + M^2_{gh}).
\eea
Because of the simple form of (3.1) we can immediately apply the general
results
obtained in ~\cite{mkg},~\cite{sigma},~\cite{josh} to evaluate the
determinants\footnote{The expression for the logarithmically divergent term
agrees with the one given in~\cite{duff}.}:
\bea
& & {i\over 2}\Tr\ln(D^2 + H_\Phi) =
\sqrt{g}\Bigg\{ {\Lambda^2\over 32\pi^2}\Tr\({1\over 6}r - H_\Phi\)
\nonumber \\
& & \qquad + \lll\Tr\({1\over 2}H_\Phi^2 - {1\over 6}rH_\Phi +
{1\over 12}G_{\mu\nu}G^{\mu\nu} +
{1\over 120}\[r^2 + 2r^{\mu\nu}r_{\mu\nu}\]\)\Bigg\}, \nonumber \\
 & & \quad \eea
and since
\begin{equation}
-{i\over 2}\Tr\ln(-i\notD + M_\Theta) = -{i\over 4}\Tr\ln[D^2 + H_\Theta],$$
$$ H_\Theta = M^2_\Theta -i[\notD,M_\Theta] +
{1\over 4}[\gamma^\mu,\gamma^\nu]G_{\mu\nu} ,\end{equation}
the fermion trace is $-{1\over 2}$ times (3.3) with the substitution
$ H_\Phi\to H_\Theta, $ and the trace includes a trace over Dirac indices, so
$$ {1\over 2}\(\Tr \;1\)_\Theta = \(\Tr \;1\)_\Phi =  2N+10. $$
Similarly, the ghost and
ghostino contributions are equivalent to, respectively, $-2$ and $+2$ times
the contribution of a four-component scalar with the masses $M_{ghost}^2 =
H_{ghost}$ and connections as determined in Section 2.  The matrix elements of
$H$ and
\begin{equation}
G_{\mu\nu} = [D_\mu,D_\nu]
\end{equation} are given in Appendix B.

The traces in (3.3--3.5) are explicitly evaluated in
Appendix B; here we simply state the result.  If $\L(g,K)$ is the standard
Lagrangian~\cite{cremmer},~\cite{kahler} for $N=1$ supergravity coupled to
matter with space-time metric $g_{\mu\nu}$ and K\"ahler potential $K$, then
the logarithmically divergent part of the one loop corrected Lagrangian is
\bea
\L_{eff} &=& \L\(g_R,K_R\) \nonumber \\ & & +
\sqrt{g}\lll\Bigg\{e^{-2K}\Bigg[ A_i\A^kR^{m\;i}_{\;n\;k}R^{n\;p}_{\;m\;q}
A_p\A^q -4R^{m\;i}_{\;n\;k} A_i\A^kA_m\A^n \nonumber \\ & &
- {2\over 3}R^m_nA_m\A^nA_j\A^j +
(R^{j\;k}_{\;n\;i}A_{jk}\A^nA\A^i + {\rm h.c.})
- R^{j\;k}_{\;\ell\;i}A_{jk}\A^{\ell n}A_n\A^i \nonumber \\ & &
- (D^{\ell}R^{j\;k}_{\;n\;i})A_{jk}\A^nA_{\ell}\A^i
- R^{j\;k}_{\;n\;i}R^{\ell\;m}_{\;j\; k}A_{\ell}\A^nA_m\A^i
- R^{\ell\;m}_{\;j\; k}A_{i\ell}\A^{jk}A_m\A^i \nonumber \\ & &
- (D_iR^{\ell\;m}_{\;j\; k})A_{\ell}\A^{jk}A_m\A^i) \Bigg]  \nonumber \\
& & +  8\hV^2 + {2\over 3}\(N + 5\)\hV M_\psi^2 +
(N + 5)M_\psi^4  \nonumber \\ & &
+ \D_\mu z^i\D^\mu\z^{\m}\Bigg(e^{-K}\Bigg[- {2\over 3}K_{i\m}
R^k_nA_k\A^n + 2R^k_{i\m j}R^{\ell\;j}_{\;n\; k}A_{\ell}\A^n
-4R^k_{i\m j}A_k\A^j \nonumber \\ & &
- R^{j\;k}_{\;n\;i}A_{jk}\A^n_{\m} -
(D_{\m}R^{j\;k}_{\;n\;i})A_{jk}\A^n - R^{\ell}_{j\m k}A_{i\ell}\A^{jk}
- (D_iR^{\ell}_{j\m k})A_{\ell}\A^{jk}\Bigg] \nonumber \\ & &
+ \[{1\over 3}(N + 29)\hV + {2\over 3}(N + 5)M^2_\psi\] K_{i\m}
- 2R_{i\m}\[{1\over 3}\hV + M^2_\psi \] \Bigg) \nonumber \\ & &
+ e^{-K}\(\D_\mu z^i\D^\mu z^j[ A_{ik\ell}\A^{\n}R^{k\;\ell}_{\;n\;j}
- R^{k\;\ell}_{\;j\; i}(A_{mk\ell}\A^m -A_{k\ell}\A)
] + {\rm h.c.}\) \nonumber \\ & &
- 4\(\D_\mu\z^{\m}\D^\mu z^iK_{i\m}\)^2 + \({N\over 6}
+ 7\)\D_\mu z^j\D^\mu z^i\D_\nu\z^{\m}\D^\nu\z^{\n}K_{i\n}K_{j\m}
\nonumber \\ & &
+ {32\over 3}\D_\mu\z^{\m}\D^\mu z^i\D_\nu\z^{\n}\D^\nu z^jK_{i\n}K_{j\m}
- {2\over 3}\D_\rho z^i\D^\rho\z^{\m}K_{i\m}\D^\mu z^j\D_\mu\z^{\n}R_{j\n}
\nonumber \\ & &
+ \D_\mu z^j\D^\mu\z^{\m}R^k_{j\m i}
\D_\nu z^{\ell}\D^\nu\z^{\n}R^i_{\ell\n k}
+ \D_\mu z^j\D^\mu z^iR^{k\;\;\ell}_{\;\;j\;\; i}
\D_\nu\z^{\n}\D^\nu\z^{\m}R_{\n k\m\ell} \nonumber \\ & &
+ 4\D_\mu\z^{\m}\D^\mu\z^{\n}\D_\nu z^j\D^\nu z^iR_{\n j\m i}
- 4\D_\mu\z^{\m}\D^\mu z^i\D_\nu z^j\D^\nu\z^{\n}R_{\m j\n i} \nonumber \\ & &
+ {1\over 2}\[\D_\mu z^j\D_\nu\z^{\m}R^k_{i\m j}
\D^\mu z^{\ell}\D^\nu\z^{\n}R^i_{k\n \ell}
- \D_\mu z^j\D_\nu\z^{\m}R^k_{i\m j}
\D^\nu z^{\ell}\D^\mu\z^{\n}R^i_{k\n \ell}\] \nonumber \\ & &
+ 4\(\L_i\A^iAe^{-K} +{\rm h.c.}\) \Bigg\}.
\eea
The classical Lagrangian $\L(g,K)$ is given in Appendix A.
Since we are neglecting gauge couplings, the gauge covariant derivative
$\D_\mu$ is here an ordinary derivative: $\D_\mu\to\pp_\mu$.
The renormalized K\"ahler potential is
\begin{equation}
K_R= K + \lll e^{-K}\[A_{ij}\A^{ij} -2A_i\A^i - 4A\A\], $$
$$ A = e^KW = (\A)^{\dag}, \;\;\;\; A_i = D_iA, \;\;\;\; \A^i =
K^{i\m}D_{\m}\A \;\;\;\; {\it etc.},
\end{equation}
and the field redefinition covariant derivative $D_i$ is defined in Appendix A.
The renormalized space-time metric is given by
\bea
g_{\mu\nu} &=& (1 - \epsilon)g^R_{\mu\nu} + \epsilon_{\mu\nu}, \nonumber \\
\epsilon &=& -\lll\Bigg[ e^{-K}\(A_{ki}\A^{ik} - {1\over 3}
R^k_nA_k\A^n\) + {N+17\over 2}\hV + {2N+16\over 3}M^2_\psi +
{2\over 3}r \Bigg], \nonumber \\
\epsilon_{\mu\nu} & = & \lll\bigg\{\[{N-19\over 6}
g_{\mu\nu}\D_\rho z^i\D^\rho\z^{\m} - {N+29\over 6}\(
\D_\mu z^i\D_\nu\z^{\m} + \D_\nu z^i\D_\mu\z^{\m} \) \]K_{i\m} \nonumber
\\ & & \qquad
- {2\over 3}R_{i\m}g_{\mu\nu}\D_\rho z^i\D^\rho\z^{\m}
+ {1\over 6}(N + 5)r_{\mu\nu} \bigg\}.
\eea
The term in (3.6) proportional to $\L_I$ can be removed by a (nonholomorphic)
scalar field redefinition:
$$ z^i\to z^i + X^i,\;\;\;\; \L(z)\to\L(z) + X^i\L_i, \;\;\;\; X^i = -
{\ln\Lambda^2\over 8\pi^2}\A^iAe^{-K} . $$

The quadratically divergent contributions
are given by (B.20--B.21).
However, as emphasized in~\cite{josh},
the relative coefficients of the quadratically divergent terms are
unreliable as they depend on the explicit regularization procedure
used~\cite{sigma},~\cite{casimir}.  Therefore we have actually only identified
all the ultraviolet divergent terms
at one loop in the effective bosonic Lagrangian of supergravity theories, and
determined the coefficients of the logarithmically divergent terms.
The full quadratically divergent one loop correction to the effective
Lagrangian for a toy model~\cite{wit} has been determined~\cite{casimir} for
the
leading term in the number $N_{ns}$ of gauge nonsinglet chiral multiplets, for
which
the definition of a Pauli-Villars regularization scheme consistent with the
requisite symmetries is straightforward;  the effective cut-off in that scheme
coincides with the one required~\cite{tom} for consistency, within a
supersymmetric theory, between the chiral and conformal anomalies under modular
transformations in target space; the conformal anomaly is related to the choice
of cut-off, while the axial anomaly is finite and unambiguously determined.
Defining consistent regularization schemes for
higher spin loops appears much more problematic.  Moreover, in realistic
theories
the effective cut-offs appearing in different terms will not even have a
uniform
dependence on the scalar fields.  The issue of removing the breaking of modular
invariance induced by the quadratically divergent terms has yet to be
addressed.

\section{String Models}
\setcounter{equation}{0}\indent

We have shown that most of the K\"ahler curvature independent terms that appear
in the logarithmically divergent one-loop contributions to the effective
supergravity action can be absorbed into field redefinitions or interpreted as
a
renormalization of the K\"ahler potential of the standard classical Lagrangian.
The curvature dependent terms vanish for models with a minimal kinetic term:
$K_{i\m} = \delta_{im}.$  More interesting for string phenomenology is
a class of theories in which the K\"ahler potential separates into
disconnected sectors that depend on different subsets $\alpha$ of chiral
fields:
\begin{equation}
K = \sum K^\alpha, \;\;\;\; \pp_i\pp_jK^\alpha = n_\alpha K^\alpha_iK^\alpha_j,
$$
$$ R^\alpha_{\n j\m i} = -n_\alpha(K^\alpha_{i\n}K^\alpha_{j\m} +
K^\alpha_{j\n}K^\alpha_{i\m}), \;\;\;\; D_iR_{\n j\m k} = 0,
\end{equation}
because the metric is covariantly constant.  These results apply to Witten's
toy
model~\cite{wit} and to the untwisted sector of orbifold
compactifications~\cite{orb}.  In models
with three matter generations in the untwisted sector, there is further
simplification because~\cite{orb},~\cite{dkl} $\pp_i\pp_mW = 0$ if
$K_{i\m} \ne 0$ and also $n_\alpha=+1$ for all $\alpha$.
This is true for the three matter + moduli generations, as
well as for the dilaton, which (neglecting nonperturbative effects) has no
superpotential.  In this case one finds for the covariant derivatives:
\begin{equation}
A_{ij} = 0 \;\;\; {\rm if} \;\;\; \alpha_i = \alpha_j.
\end{equation}
(Here the notation $\alpha_i=\alpha_j$ means ``if $i$ and $j$ belong
to the same subset''.)  Then since
\begin{equation}
R^{i\;\;j}_{\;m\;\;n} = 0 \;\;\;{\rm if} \;\;\; \alpha_i \ne \alpha_j,
\end{equation}
the result (3.6) reduces to
\bea
\L_{eff} &=& \L\(g_R,K_R\) \nonumber \\ & & +
\sqrt{g}\lll\Bigg\{e^{-2K}\[\sum_\alpha\(N + 7\)(A_i\A^i)^2_\alpha -
{2\over 3}A_i\A^i\sum_\alpha N_\alpha(A_j\A^j)_\alpha\] \nonumber \\ & &
%+ 2\({1\over 3}V + M_\psi^2\)e^{-K}\sum_\alpha\(A_i\A^i\)_\alpha
+2 \hV^2 + {2\over 3}\(N -1\)\hV M_\psi^2 +
(N -1 )M_\psi^4  \nonumber \\ & &
+ \D_\mu z^i\D^\mu\z^{\m}e^{-K}\[-{2\over 3}K_{i\m}
\sum_\alpha N_\alpha(A_j\A^j)_\alpha + \sum_\alpha\(2N_\alpha + 8\)
K^\alpha_{i\m}(A_j\A^j)_\alpha\] \nonumber \\ & &
+6e^{-K}\sum_\alpha\(\D_\mu z^iA^i\)_\alpha\(\D^\mu\z^{\m}\A_{\m}\)_\alpha
\nonumber \\ & &
+ \D_\mu z^i\D^\mu\z^{\m}e^{-K}\[\({N + 27\over 3}\hV + {2\over 3}(N + 2)
M^2_\psi\) K_{i\m} - 2R_{i\m}\({1\over 3}\hV + M^2_\psi \)\] \nonumber \\
& & - 4\(\D_\mu\z^{\m}\D^\mu z^iK_{i\m}\)^2 + \({N\over 6}
+ 7\)\D_\mu z^j\D^\mu z^i\D_\nu\z^{\m}\D^\nu\z^{\n}K_{i\n}K_{j\m}
\nonumber \\ & &
+ {32\over 3}\D_\mu\z^{\m}\D^\mu z^i\D_\nu\z^{\n}\D^\nu z^jK_{i\n}K_{j\m}
+ {2\over 3}\D_\rho z^i\D^\rho\z^{\m}K_{i\m}\sum_\alpha\(N_\alpha+1\)\D^\mu
z^j\D_\mu\z^{\n}K^\alpha_{j\n} \nonumber \\ & &
+ \sum_\alpha\bigg[\(N_\alpha + {11\over 2}\)\(
\D_\mu z^i\D^\mu\z^{\m}K^\alpha_{i\m}\)^2 +
{1\over 2}\(N_\alpha - 5\)\D_\mu z^j\D^\mu z^i\D_\nu\z^{\n}\D^\nu\z^{\m}
K^\alpha_{i\m}K^\alpha_{j\n} \nonumber \\ & &
- {1\over 2}\(N_\alpha - 8\)\D_\mu\z^{\m}\D^\mu z^i\D_\nu z^j\D^\nu\z^{\n}
K^\alpha_{j\m}K^\alpha_{i\n}\bigg]\Bigg\},
\eea
where
$$N= \sum_\alpha N_\alpha, \;\;\;\; (A_iB^i)_\alpha = \sum_{i\in\alpha}
A_iB^i, $$ and for the dilaton $s$
\begin{equation}
N_s = 1, \;\;\;\; A_s\A^s = A\A.
\end{equation}

Our results will be extended to include the gauge sector in~\cite{future}.

\vskip .3in
\noindent{\bf Acknowledgements.}  We thank Aris Papadopoulos for helpful
discussions, and MKG would like to acknowledge the hospitality
of the Institute for
Theoretical Physics, University of California at Santa Barbara, where much of
this work was done.  This work  was  supported  in  part by the
Director, Office of Energy Research, Office of High Energy and Nuclear Physics,
Division of High Energy Physics of the U.S. Department of Energy under Contract
DE-AC03-76SF00098 and in  part by the National  Science Foundation under grant
PHY-90-21139.

\vskip .3in
\appendix
\noindent{\large \bf Appendix}
\def\ksubsection{\Alph{subsection}}
\def\theequation{\ksubsection.\arabic{equation}}

%       reset section commands

\catcode`\@=11

\def\thesubsection{\Alph{subsection}.}
\def\thesubsubsection{\arabic{subsubsection}.}

\subsection{Conventions and Notation}
\setcounter{equation}{0}

Our Dirac matrices and space-time metric signature $(+ - - -)$ are
those of Bjorken and Drell or Itzykson and Zuber. We use uppercase notation
($R,\;\Gamma$) for derivatives of the K\"ahler metric, and lowercase
($r,\;\gamma$) for those of the space-time metric.  Our sign conventions for,
respectively, the Riemann tensor, Ricci tensor and curvature scalar are as
follows:
\begin{equation}
r^\mu_{\nu\rho\sigma} = g^{\mu\lambda}r_{\lambda\nu\rho\sigma} =
\pp_\sigma\gamma^\mu_{\nu\rho} -
\pp_\rho\gamma^\mu_{\nu\sigma} + \gamma^\mu_{\sigma\lambda}
\gamma^\lambda_{\nu\rho} -
\gamma^\mu_{\rho\lambda}\gamma^\lambda_{\nu\sigma},$$
$$ r_{\mu\nu} = r^\rho_{\mu\rho\nu}, \;\;\;\; r = g^{\mu\nu}r_{\mu\nu},
\end{equation}
and covariant differentiation is defined by
\begin{equation}
\nabla_\mu A_\nu = \pp_\mu A_\nu - \gamma^\rho_{\mu\nu} A_\rho, \;\;\;\;
{\it etc.} \end{equation}
The scalar field redefinition covariant quantities are defined identically with
\begin{equation}
g_{\mu\nu}\to Z_{IJ}, \;\;\;\; \gamma\to \Gamma, \;\;\;\; r\to R, \;\;\;\;
\nabla_\mu \to D_I \;\;\;\; I = i,\ibar, \end{equation}
where $z^i,\;\z^{\m} = (z^m)^{\dag}$ are the scalar partners of left and right
handed Weyl fermions, respectively.
Because the scalar metric is K\"ahler, there is only one type of nonvanishing
element of the Riemann tensor, namely
\bea  & &
R^i_{jk\m} = \pp_{\m}\Gamma^i_{jk} = D_{\m}\Gamma^i_{jk}=-R^i_{j\m k},
\nonumber \\ & &
R_{\n jk\m} = R_{\n kj\m} = R_{\m jk\n} = R_{\m kj\n} \nonumber \\ & &
\quad = -R_{\n j\m k} = -R_{\n k\m j}=-R_{\m j\n k}=-R_{\m k\n j}.
\eea
Note that since $R^i_{jk\ell}=0, \; [D_i,D_j]=0$, and the tensors
\begin{equation}
A_{i_1\cdots i_n} = D_{i_1}\cdots D_{i_n}A, \;\;\;\; \A^{i_1\cdots i_n} =
D^{i_1}\cdots D^{i_n}\A, \end{equation}
are symmetric in all indices.  It follows  from the Bianchi identities that
$D_iR^n_{j\m k}$ is totally symmetric in \{$ijk$\}.

We work in the K\"ahler covariant formalism~\cite{kahler}, which differs from
that of Cremmer {\it et al.}~\cite{cremmer} by a phase transformation on the
fermions that removes  phases proportional to Im$\(W/\W\)$, where $W$ is the
superpotential.  In this formalism the fermion U(1) K\"ahler connection is just
\begin{equation}
\Gamma_\mu = {i\over 4}\(K_i\D_\mu z^i -K_{\m}\D_\mu\z^{\m}\),
\end{equation}
where $\D_\mu$ is the gauge covariant derivative.
It is convenient to introduce the notation
\begin{equation}
A = e^KW, \;\;\;\; \A = e^K\W.
\end{equation}
Then the classical potential is $V = \hV + \D$, where
\begin{equation}
\hV = e^{-K}(A_i\A^i - 3A\A), \;\;\;\;
\D = {1\over 2x}\D_a\D^a, \;\;\;\; \D_a = K_i(T^az^i).\end{equation}
With these conventions the tree level
Lagrangian~\cite{cremmer},~\cite{kahler} for the case
$f(z)_{ab} = \delta_{ab}f(z) = \delta_{ab}
[x(z,\z) + iy(z,\z)]$ is
\bea
g^{-{1\over 2}}\L(g,K,f) &=& {1\over 2}r + K_{i\m}\D^\mu z^i\D_\mu\z^{\m}
- {x\over 4}F_{\mu\nu}F^{\mu\nu} -
{y\over 4}g^{-{1\over 2}}{\tilde F}_{\mu\nu}F^{\mu\nu} - V \nonumber \\
& & + {ix\over 2}\bl\notD\lambda + iK^{i\m}\(\bc_L^{\m}\notD\chi_L^i +
\bc_R^i\notD\chi_R^{\m}\) \nonumber \\ & &
+ e^{-K/2}\({1\over 4}f_i\A^i\bl_R\lambda_L - A_{ij}\bc^i_R\chi^j_L
+ {\rm h.c.}\) \nonumber \\ & &
+ \(i\bl^a_R\[2K_{i\m}(T_a\z)^{\m} - {1\over 2x}f_i\D_a +
{1\over 4}\sigma_{\mu\nu}F^{\mu\nu}_af_i\]\chi^i_L + {\rm h.c.}\)
\nonumber \\ & &
+ \L_\psi + {\rm four-fermion \; terms},
\eea
with $\L_\psi$ given in (2.7).
In the notation of~\cite{josh} [see eq.(3.91)], the masses operating on the
left-handed gravitino and chiral fermions are
\begin{equation}
m^\psi = e^{-K/2}\A, \;\;\;\; m^\chi_{ij} = 2e^{-K/2}A_{ij}.
\end{equation}
These are related to the elements of $M_\Theta$ in (3.1) by
(see~\cite{josh})
\begin{equation}
M^\mu_\nu = g^\mu_\nu e^{-K/2}\A = g^\mu_\nu M_\psi, \;\;\;\;
M^{\m}_j = K^{\m i}e^{-K/2}A_{ij} = e^{-K/2}A^{\m}_j.
\end{equation}
Note that the normalization of our chiral fermions is the same as
in~\cite{cremmer}, which differs by a factor $\sqrt{2}$ from~\cite{kahler}.
The covariant derivatives $D_\mu$ include the spin connection, the gauge
connection, the K\"ahler connection (A.6), the affine connection,
and the field reparameterization
connection for chiral fields. For fermions:
\bea
D_\mu\psi &=& \[\nabla_\mu + {1\over 4}\gamma_\nu\(\nabla_\mu\gamma^\nu\)
+ i\gamma_5\Gamma_\mu\]\psi, \nonumber \\
D_\mu\chi^I & =& \[\D_\mu + {1\over 4}\gamma_\nu\(\nabla_\mu\gamma^\nu\)
- i\gamma_5\Gamma_\mu\]\chi^I + \D_\mu z^J\Gamma^I_{JK}\chi^K.\eea
(The gauginos have the same K\"ahler weight as the gravitino, and an additional
connection which is given in (C.7) below.) Operating on a function of
scalar fields, $D_\mu=\D_\mu z^I D_I$, where $\D_\mu$ is gauge and
general coordinate covariant.

\subsection{Operators and Traces}
\setcounter{equation}{0}

In this Appendix we list the matrix elements of the operators appearing in Eqs.
(3.1--3.5) and the traces needed to evaluate the divergent contributions to the
one-loop effective action (3.2).  We drop all total derivatives in the traces.

\subsubsection{The bosonic sector}

In the absence of the Yang-Mills sector, the operator $H_\Theta$ can be
expressed as~\cite{tini},~\cite{vid},~\cite{mkg},~\cite{sigma},~\cite{josh}
\begin{equation}
Z_\Theta H_\Theta = H + X + Y, \end{equation}
with
\begin{equation}
Z_{i\mu\nu} = 0, \;\;\;\; Z_{i\m} = K_{i\m}, \;\;\;\; Z_{ij}= Z_{\m\n} =0,$$
$$ Z_{\mu\nu,\rho\sigma} = P_{\mu\nu,\rho\sigma} = {1\over
2}\(g_{\mu\rho}g_{\nu\sigma} + g_{\nu\rho}g_{\mu\sigma} -
g_{\mu\nu}g_{\rho\sigma}\) = {1\over 16}P^{-1}_{\mu\nu,\rho\sigma}.
\end{equation}
The nonvanishing elements of $Z_\Theta H_\Theta$
are $H_{IJ},\;X_{\mu\nu}$ and $Y_{\mu I}$, with
\begin{equation}
H_{IJ} = V_{IJ} + h_{IJ}, \;\;\;\; h_{IJ}= \R_{IJ} + U_{IJ}, $$
$$ V_{IJ} = D_ID_JV, \;\;\;\; \R_{IJ} = \D_\mu z^K\D_\nu z^L R_{IKLJ}, \;\;\;\;
U_{IJ} = - 2\D_\mu z^K\D_\nu\z^L Z_{IK}Z_{JL} $$
$$ X_{\mu\nu,\rho\sigma} = - 2P_{\mu\nu,\rho\sigma}V
- {1\over 4}\[r_{\mu\rho\nu\sigma} + r_{\nu\rho\mu\sigma} \], $$
$$ Y_{\mu\nu I} = Y_{I\mu\nu} = - Z_{IJ}D_\mu\D_\nu z^J.
\end{equation}
The contribution $U_{IJ}$ to the scalar ``squared mass'' arises from the
graviton gauge fixing term which is the same as in
Refs.~\cite{tini},~\cite{vid},~\cite{josh},~\cite{lahanas}.  The expressions
for $X$ and $Y$ are simpler than in those references because of the propagator
modification introduced in Section 2.  Using
\begin{equation}
\hV_i = e^{-K}[A_{ji}\A^j - 2A_i\A], \;\;\;\; \hV_{ij} = e^{-K}[A_{jik}\A^k -
A_{ij}\A], $$
$$ \hV_i^j = K^{j\m}\hV_{i\m} = e^{-K}[A_{ki}\A^{jk} - A_i\A^j +
\delta_i^jA_k\A^k - 2\delta_i^jA\A + R^{k\;j}_{\;n\;i}A_k\A^n],
\end{equation}
we obtain the following traces (Lorentz indices are raised with $g^{\mu\nu}$
and
scalar indices are raised with $K^{i\m}$):
\bea
\Tr H &=& 2H_i^i = 2e^{-K}[A_{ki}\A^{ik} - R^k_nA_k\A^n]
+ 2(N-1)\hV + 2(N-3)M^2_\psi \nonumber \\ & &
- \D^\mu z^i\D_\mu\z^{\m}(2R_{i\m} + 4K_{i\m}), \nonumber \\
{1\over 2}\Tr H^2 &=& H^i_jH_i^j  + H_{ij}H^{ij}  \nonumber \\
&=& e^{-2K}[A_{ij}\A^{jk}\A^{im}A_{mk} - 2A_{ij}\A^{jk}\A^iA_k
+A_{ij}\A^{ij}(2\A^kA_k - 3A\A)  \nonumber \\ & &
+ 2A_{ij}\A^{jk}R^{m\;i}_{\;n\;k}A_m\A^n
+ A_i\A^kR^{m\;i}_{\;n\;k}R^{n\;p}_{\;m\;q}A_p\A^q -2R^{m\;i}_{\;n\;k}
A_i\A^kA_m\A^n \nonumber \\ & &
- 2R^m_nA_m\A^n\(\hV + M_\psi^2\) + (N-1)\hV^2 + 2(N-1)\hV M_\psi^2
+ (N + 3)M_\psi^4 \nonumber \\ & &
+ A_{kij}\A^{ijm}\A^kA_m - (A_{ijk}\A^{ik}\A^j A + {\rm h.c.})]
\nonumber \\ & &
-2(\D_\mu z^j\D^\mu z^ie^{-K}[A_{jik}\A^k -A_{ij}\A] +{\rm h.c.})
\nonumber \\ & &
- \D_\mu\z^{\m}\D^\mu z^i\lbr 4e^{-K}[A_{ij}\A^j_{\m} -A_i\A_{\m}] +
\(4K_{i\m} + 2R_{i\m}\)\(\hV + M_\psi^2\)\rbr \nonumber \\ & &
+ 2\D_\mu z^j\D^\mu\z^{\m}R^k_{j\m i}e^{-K}[A_{k\ell}\A^{\ell i} -3A_k\A^i +
+ R^{\ell\;i}_{\;n\; k}A_{\ell}\A^n] \nonumber \\ & &
- (\D_\mu z^j\D^\mu z^iR^{k\;\ell}_{\;j\; i}e^{-K}[A_{mk\ell}\A^m
-A_{k\ell}\A] +{\rm h.c.}) \nonumber \\ & &
+ \D_\mu z^j\D^\mu\z^{\m}R^k_{j\m i}
\D_\nu z^{\ell}\D^\nu\z^{\n}R^i_{\ell\n k}
+ \D_\mu z^j\D^\mu z^iR^{k\;\;\ell}_{\;\;j\;\; i}
\D_\nu\z^{\n}\D^\nu\z^{\m}R_{\n k\m\ell} \nonumber \\ & &
+ 4\D_\mu z^j\D^\mu z^i\D_\nu\z^{\m}\D^\nu\z^{\n}K_{i\n}K_{j\m}
+ 4\D_\mu\z^{\m}\D^\mu z^i\D_\nu\z^{\n}\D^\nu z^jK_{i\n}K_{j\m} \nonumber
\\ & &
+ 4\D_\mu\z^{\m}\D^\mu\z^{\n}\D_\nu z^j\D^\nu z^iR_{\n j\m i}
- 4\D_\mu\z^{\m}\D^\mu z^i\D_\nu z^j\D^\nu\z^{\n}R_{\m j\n i},
\eea
\bea
\Tr X &=& -20V + 2r, \nonumber \\
\Tr X^2 &=& 40V^2 - 8rV + 8r_{\mu\nu}r^{\mu\nu} - 2r^2
+ {\rm total\; derivative},\eea
and
\begin{equation}
\Tr Y^2 = 4\(D_\mu\D_\nu z^i\)\(D^\mu\D^\nu z^{\m}\)K_{i\m} +
4r^{\mu\nu}\D_\mu z^i\D_\nu z^{\m}K_{i\m}.
\end{equation}
Finally we need
\bea
G_{\mu\nu} &=& \(G_z + G_G\)_{\mu\nu}, \nonumber \\
\(G^z_{\mu\nu}\)^I_J &=& \(R_{\mu\nu}\)^I_J =
\D^\mu z^I\D_\nu z^L R^I_{JLK},\nonumber \\
 \Tr R_{\mu\nu}R^{\mu\nu} &=& 2\[\D_\mu z^j\D_\nu\z^{\m}R^k_{i\m j}
\D^\mu z^{\ell}\D^\nu\z^{\n}R^i_{k\n \ell}
- \D_\mu z^j\D_\nu\z^{\m}R^k_{i\m j}
\D^\nu z^{\ell}\D^\mu\z^{\n}R^i_{k\n \ell}\], \nonumber \\
 \(G^G_{\mu\nu}\)_{\alpha\beta,\gamma\delta}
 & =& \delta_{\alpha\beta,\rho\sigma}\(r^\rho_{\gamma\mu\nu}
g^\sigma_\delta + r^\rho_{\delta\mu\nu}g^\sigma_\gamma\), \nonumber \\
 \Tr \[ G_{\mu\nu}G^{\mu\nu}\]_{G}    &=&
-6r^{\rho\sigma\mu\nu}r_{\rho\sigma\mu\nu} =
12\({1\over 2}r^2 - 2r^{\mu\nu}r_{\mu\nu}\) + {\rm total\; derivative}.
\eea

\subsubsection{The fermion sector}

The metric is
\begin{equation}
Z_{\chi^I\chi^J} = 2Z_{IJ}, \;\;\;\; Z_{\alpha\alpha} = -2, \;\;\;\;
Z_{\mu\nu} = - g_{\mu\nu}, \;\;\;\; Z_{\chi\alpha}=Z_{\chi\mu}=Z_{\alpha\mu}=0.
\end{equation}
The matrix elements of $M_\Theta$ are given by (2.16), (2.17), (A.11) and
\begin{equation}
M^\mu_I = 2Z_{IJ}\D^\mu z^J, \;\;\;\; M^I_\mu = - \D_\mu z^I.
\end{equation}
We also need their
covariant derivatives, which have been defined in~\cite{josh}, with the
difference that the K\"ahler connection is here given by (A.6).  In evaluating
these derivatives it is useful to recall that the gaugino has opposite K\"ahler
weight from the chiral fermions and the auxiliary field $\alpha$.  One finds
for the covariant derivatives of the matrix elements defined in (2.16), (2.17),
and (A.7):
\begin{equation}
D_\mu \mu_{ij} = e^{-K/2}\(A_{ijk}\D_\mu z^k + \D_\mu\z^{\m}[A_iK_{j\m} +
A_jK_{i\m} + A_nR^n_{i\m j}]\), $$
$$ D_\mu m_i = e^{-K/2}\(A_{ik}\D_\mu z^k + \D_\mu\z^{\m}AK_{i\m}\), \;\;\;\;
D_\mu M = e^{-K/2}A_k\D_\mu z^k .
\end{equation}
The nonvanishing matrix elements of $G_{\mu\nu}$ are
\bea
\(G_{\mu\nu}\)^I_J &=& \(R_{\mu\nu}\)^I_J + \delta^I_J\(- \gamma_5
\Gamma_{\mu\nu} + Z_{\mu\nu}\), \nonumber \\
\(G_{\mu\nu}\)^\alpha_\alpha &=& -\gamma_5\Gamma_{\mu\nu} + Z_{\mu\nu},
\nonumber \\
\(G_{\mu\nu}\)^\rho_\sigma &=& \delta^\rho_\sigma\(\gamma_5\Gamma_{\mu\nu} +
Z_{\mu\nu}\) - {1\over 2}r^\rho_{\sigma\nu\mu},
\eea
where
\begin{equation}
\Gamma_{\mu\nu} = {1\over 2}\(\D_\nu\z^{\m}K_{i\m}\D_\mu z^i
-\D_\mu\z^{\m}K_{i\m}\D_\nu z^i\), $$
$$ Z_{\mu\nu} = {1\over 4}r_{\rho\sigma\nu\mu}\gamma^\rho\gamma^\sigma.
\end{equation}
Then we find the following traces:
\bea
{1\over 4}\Tr M_\Theta^2 &=& e^{-K}\[A_{ij}\A^{ij} - 2\hV
+ 2M^2_\psi \] - 4\D^\mu z^i\D_\mu\z^{\m} K_{i\m},\nonumber \\
{1\over 4}\Tr M_\Theta^4 &=& e^{-2K}[A_{ij}\A^{jk}\A^{im}A_{km}
- 4A_{ij}\A^{jk}A_k\A^i \nonumber \\ & &
+ 4(A\A^iA_{ik}\A^k +{\rm h.c.})] + 2\hV^2 - 4\hV M_\psi^2 - 10M^4_\psi
\nonumber \\ & &
-8e^{-K}\D_\mu z^i\D^\mu\z^{\m}\(A_{ij}\A^j_{\m} + K_{i\m}A\A -A_i\A_{\m}\)
\nonumber \\ & &
 -4e^{-K}(\D_\mu z^i\D^\mu z^jA_{ij}\A + {\rm h.c.}) +
8\D_\mu z^j\D^\nu z^i\D_\nu\z^{\m}\D^\mu\z^{\n}K_{i\n}K_{j\m}, \nonumber \\
 {1\over 4}\Tr|D_\mu M_\Theta|^2  & = &
e^{-K}\D_\mu z^i\D^\mu\z^{\m}[A_{ijk}\A^{jk}_{\m}
+ 10A_i\A_{\m} \nonumber \\ & &
+ 4R_{i\m j}^kA_k\A^j + R_{j\m\;n}^{\ell}R^{j\;n}_{\;k\;i}A_{\ell}\A^k
 -2\A^j_{\m}A_{ij} ] + 2K_{i\m}\(\hV + 2M_\psi^2\) \nonumber \\ & &
+ e^{-K}\{\D_\mu z^i\D^\mu z^j
[2A_{ijk}\A^k + A_{ik\ell}\A^nR^{k\;\ell}_{\;n\;j} -2\A A_{ij}] + {\rm h.c.}\}
\nonumber \\ & &
- 4\(D_\mu\D_\nu z^j\)\(D^\mu\D^\nu\z^{\m}\)K_{j\m}, \nonumber \\
{1\over 4}\Tr\([\gamma^\mu,\gamma^\nu]G_{\mu\nu}\) &=& (N + 5)r,
\nonumber \\
{1\over 4}\Tr\(M^2_\Theta[\gamma^\mu,\gamma^\nu]G_{\mu\nu}\) &=&
{1\over 4}r\Tr M^2_\Theta, \nonumber \\
{1\over 16}\Tr\([\gamma^\mu,\gamma^\nu]G_{\mu\nu}\)^2 &=& - 2(N + 5)
\Gamma_{\mu\nu}\Gamma^{\mu\nu} - \[\Tr G_{\mu\nu}G^{\mu\nu}\]_z
\nonumber \\ & &
+ {1\over 4}(N-3)r^2 + 8r_{\mu\nu}r^{\mu\nu}, \nonumber \\
\Tr G_{\mu\nu}G^{\mu\nu} &=& 4(N + 5)\Gamma_{\mu\nu}\Gamma^{\mu\nu} +
2\[\Tr G_{\mu\nu}G^{\mu\nu}\]_z \nonumber \\ & &
+ (N+13)\({1\over 2}r^2 - 2r_{\mu\nu}r_{\mu\nu}\),
\eea
where
\begin{equation}
\Gamma_{\mu\nu}\Gamma^{\mu\nu} = {1\over 2}
\(\D_\mu z^j\D^\mu z^i\D_\nu\z^{\m}\D^\nu\z^{\n}K_{i\n}K_{j\m}
- \D_\mu\z^{\m}\D^\mu z^i\D_\nu\z^{\n}\D^\nu z^jK_{i\n}K_{j\m}\).
\end{equation}

\subsubsection{The ghost sector}
For the graviton ghost we have~\cite{tini},~\cite{vid}
\bea & &
H_{gh}^{\mu\nu} = -2\D^\mu z^I\D^\nu z^JZ_{IJ} - r^{\mu\nu}, \;\;\;\;
(G^{gh}_{\mu\nu})^\alpha_\beta =
r^\alpha_{\beta\mu\nu} \nonumber \\ & &
\Tr H_{gh} = -4\D_\mu z^i\D^\mu\z^{m}K_{i\m} - r \nonumber \\ & &
\Tr H_{gh}^2 = 8\D_\mu z^j\D^\mu z^i\D_\nu\z^{\m}\D^\nu\z^{\n}K_{i\n}
K_{j\m} + 8\D_\mu\z^{\m}\D^\mu z^i\D_\nu\z^{\n}\D^\nu z^jK_{i\n}K_{j\m}
\nonumber \\ & & \qquad
+ 8r_{\mu\nu}\D^\mu z^i\D^\nu\z^{\m}K_{i\m} + r_{\mu\nu}r^{\mu\nu},
\nonumber \\ & &
\Tr G_{\mu\nu}G^{\mu\nu} = -r_{\alpha\beta\mu\nu}r^{\alpha\beta\mu\nu}
= r^2 - 4r_{\mu\nu}r^{\mu\nu} + {\rm total\; derivative} .
\eea

\noindent
For the gravitino ghost, $H_{Gh} = M^2_{Gh}$ is given by (2.12), and
\begin{equation}
[D_\mu,D_\nu] = G_{\mu\nu} = \gamma_5\Gamma_{\mu\nu} + Z_{\mu\nu}.
\end{equation}
We get
\bea
\Tr H_{Gh} &=& 4\(\hV + M^2_\psi\) - 4\D_\mu z^iK_{i\m}\D^\mu\z^{\m} - r,
\nonumber \\
\Tr H_{Gh}^2 &=& 4\(\hV + M_\psi^2\)^2 + 4
\(\D_\mu z^iK_{i\m}\D^\mu\z^{\m}\)^2 + {1\over 4}r^2,\nonumber \\ & &
- \(8\D_\mu z^iK_{i\m}\D^\mu\z^{\m} + 2r\)\(\hV + M_\psi^2\)
- 4\D_\mu z^i\D^\mu\z^{\m}A_i\A_{\m}e^{-K} \nonumber \\ & &
+ 2r\D_\mu z^iK_{i\m}\D^\mu\z^{\m} - 6\Gamma_{\mu\nu}\Gamma^{\mu\nu},
\nonumber \\
\Tr G_{\mu\nu}G^{\mu\nu} &=& 4\Gamma_{\mu\nu}\Gamma^{\mu\nu}
+ {1\over 2}r^2 - 2r^{\mu\nu}r_{\mu\nu}.
\eea

\subsubsection{Supertraces}

If we define
\begin{equation}
\STr F = \Tr F_\Phi - {1\over 2}\Tr F_\Theta - 2\Tr F_{gh} + 2\Tr F_{Gh},
\end{equation}
the effective Lagrangian (3.2) is
\begin{equation}
\L_1 = {\Lambda^2\over 32\pi^2}\STr H
 + \lll\STr\({1\over 2}H^2 - {1\over 6}rH + {1\over 12}G_{\mu\nu}G^{\mu\nu}\),
\end{equation}
with
\bea
-{r\over 6}\STr H &=&
 {N+1\over 12}r^2 - {N-5\over 3}r\hV - {N-1\over 3}rM_\psi^2
+ {1\over 3}rR^i_jA_i\A^j \nonumber \\ & &
+ {1\over 3}r\D_\mu z^i\D^\mu\z^{\m}\(R_{i\m} - 2K_{i\m}\),\nonumber \\
{1\over 2}\STr H^2 &=& e^{-2K}[ 2A_{ij}\A^{jk}\A^iA_k
+ A_{ij}\A^{ij}\(2\hV + 3M_\psi^2\)  \nonumber \\ & &
 + 2A_{ij}\A^{jk}R^{m\;i}_{\;n\;k}A_m\A^n
+ A_i\A^kR^{m\;i}_{\;n\;k}R^{n\;p}_{\;m\;q}A_p\A^q - 2R^{m\;i}_{\;n\;k}
A_i\A^kA_m\A^n \nonumber \\ & &
 - 2R^m_nA_m\A^n\(\hV + M^2_\psi\) + A_{kij}\A^{ijm}\A^kA_m \nonumber \\
& &   + (N+ 21)\hV^2 + (N + 17)M_\psi^4 + 2(N + 5)\hV M^2_\psi
 \nonumber \\ & & -
(A_{ijk}\A^{ik}\A^j A + 4A\A^i\A^jA_{ij} + {\rm h.c.})] \nonumber \\ & &
 + e^{-K}(\D_\mu z^j\D^\mu z^i[4A_{ij}\A + A_{ik\ell}\A^n
R^{k\;\;\ell}_{\;\;n\;\;j}] +{\rm h.c.}) \nonumber \\ & &
 +\D_\mu\z^{\m}\D^\mu z^i\big\{e^{-K}[A_{ijk}\A^{jk}_{\m} + 2A_{ij}\A^j_{\m}
+ 2A_i\A_{\m} - 10K_{i\m}\hV \nonumber \\ & &
 + R^{\ell}_{j\m n}R^{j\;\;n}_{\;\;k\;\;i}A_{\ell}\A^k]
- 2R_{i\m}\(\hV + M^2_\psi\)\big\} \nonumber \\ & &
 + 2\D_\mu z^j\D^\mu\z^{\m}R^k_{j\m i}e^{-K}[A_{k\ell}\A^{\ell i} - A_k\A^i
+ R^{\ell\;i}_{\;n\; k}A_{\ell}\A^n] \nonumber \\ & &
 - (\D_\mu z^j\D^\mu z^iR^{k\;\ell}_{\;j\; i}e^{-K}[A_{mk\ell}\A^m
-A_{k\ell}\A] +{\rm h.c.}) \nonumber \\ & &
 + \D_\mu z^j\D^\mu\z^{\m}R^k_{j\m i}
\D_\nu z^{\ell}\D^\nu\z^{\n}R^i_{\ell\n k}
+ \D_\mu z^j\D^\mu z^iR^{k\;\;\ell}_{\;\;j\;\; i}
\D_\nu\z^{\n}\D^\nu\z^{\m}R_{\n k\m\ell} \nonumber \\ & &
 + {N + 9\over 4}\D_\mu z^j\D^\mu z^i\D_\nu\z^{\m}\D^\nu\z^{\n}K_{i\n}K_{j\m}
\nonumber \\ & &
- {N -23\over 4}\D_\mu\z^{\m}\D^\mu z^i\D_\nu\z^{\n}\D^\nu z^jK_{i\n}K_{j\m}
\nonumber \\ & &
 + 4\D_\mu\z^{\m}\D^\mu\z^{\n}\D_\nu z^j\D^\nu z^iR_{\n j\m i}
- 4\D_\mu\z^{\m}\D^\mu z^i\D_\nu z^j\D^\nu\z^{\n}R_{\m j\n i}
\nonumber \\ & &
+ {1\over 4}\Tr R_{\mu\nu}R^{\mu\nu} + 4\(\D_\mu\z^{\m}\D^\mu z^iK_{i\m}\)^2
- 5rV - 3rM_\psi^2 \nonumber \\ & &
 + 4r\D_\mu\z^{\m}\D^\mu z^iK_{i\m} -
4r^{\mu\nu}\D_\mu\z^{\m}\D_\nu z^iK_{i\m} - {N +9\over 16}r^2
+ r^{\mu\nu}r_{\mu\nu}, \nonumber \\
 {1\over 12}\STr G_{\mu\nu}G^{\mu\nu} &=& -{N+1\over 6}\Gamma_{\mu\nu}
\Gamma^{\mu\nu} + {7-N\over 48}\(r^2 - 4r^{\mu\nu}r_{\mu\nu}\).
\eea
Inserting these supertraces in (B.20) gives a contribution of the form (2.23)
with
\begin{equation}
\alpha = - {2\over 3}\lll, \;\;\;\; \beta = {N+5\over 6}\lll, $$
$$ \epsilon_0 = -\lll\lbr e^{-K}\(A_{ij}\A^{ij} -
{2\over 3}R^i_jA_i\A^j\) + {2N+20\over 3}\hV +
{2N+16\over 3}M^2_\psi\rbr, $$
$$ H_{\mu\nu} = H^1_{\mu\nu} = \lll\bigg\{
g_{\mu\nu}\D_\rho z^i\D^\rho\z^{\m}[{20\over 3}K_{i\m} + {2\over 3}R_{i\m}]
- 4\[\D_\mu z^i\D_\nu\z^{\m} K_{i\m} + \D_\nu z^i\D_\mu\z^{\m}
K_{i\m}\]\bigg\},
\end{equation}
so the metric redefinition (2.4) gives (3.8), and we get
a correction $\Delta_r\L$ given by the last term in (2.25):
\bea
{1\over\sqrt{g}}\Delta_r\L_1
 &=&  \lll\Bigg[\lbr -2e^{-K}\(A_{ki}\A^{ik} - {2\over
3}R^k_nA_k\A^n\) - (N+17)\hV - {4N+32\over 3}M^2_\psi -
{4\over 3}r\rbr \hV \nonumber \\ & &
 + \Bigg[K_{i\m}\Bigg\{ {N+59\over 3}\hV + e^{-K}\(A_{ki}\A^{ik} -
{2\over 3}R^k_nA_k\A^n\) \nonumber \\ & &
+ {2N+16\over 3}M^2_\psi + {2\over 3}r \Bigg\}
+{4\over 3}R_{i\m}\hV\Bigg]\D_\rho z^i\D^\rho\z^{\m}  \nonumber \\
& &
 - \lbr\({2\over 3}R_{i\m} + 8K_{i\m}\)\D_\rho z^i\D^\rho\z^{\m} \right.
\nonumber \\ & & \qquad \left.
+ {N+29\over 6}\(\D_\mu z^i\D_\nu\z^{\m} + \D_\nu z^i\D_\mu\z^{\m}\)K_{i\m}
\rbr\D_\mu z^j\D^\mu\z^{\n}K_{i\n} \Bigg].
\eea
The K\"aher potential redefinition (3.7) absorbs the contribution $\Delta_K\L$,
where
\bea
{1\over\sqrt{g}}\Delta_K\L_1 &=& \lll\big\{ e^{-2K}[-(A_{ijk}\A^{jk}A\A^i
+ R^{j\;k}_{\;n\;i}A_{jk}\A^nA\A^i + {\rm h.c.}) + 3A_{ij}\A^{ij}A\A
  \nonumber \\ & &
+ A_{ijk}\A^{jkn}A_n\A^i + 2A_{ij}\A^{jn}A_n\A^i \nonumber \\ & &
+ R^{j\;k}_{\;\ell\;i}A_{jk}\A^{\ell n}A_n\A^i +
(D^{\ell}R^{j\;k}_{\;n\;i})A_{jk}\A^nA_{\ell}\A^i
+ R^{j\;k}_{\;n\;i}R^{\ell\;m}_{\;j\; k}A_{\ell}\A^nA_m\A^i + \nonumber \\
& &  + 2R^{\ell\;m}_{\;j\; i}A_{\ell n}\A^{jn}A_m\A^i
+ R^{\ell\;m}_{\;j\; k}A_{i\ell}\A^{jk}A_m\A^i +
(D_iR^{\ell\;m}_{\;j\; k})A_{\ell}\A^{jk}A_m\A^i \nonumber \\ & &
 + 2R^{k\;\;j}_{\;\;n\;\;i}A_k\A^nA_j\A^i] - 4\hV^2 -20M^2_\psi\hV
- 36M_\psi^4 \nonumber \\ & &
 + e^{-K}[A_{ijk}\A^{jk}_{\m}
+ 2A_{ij}\A^j_{\m} + R^{j\;k}_{\;n\;i}A_{jk}\A^n_{\m} +
(D_{\m}R^{j\;k}_{\;n\;i})A_{jk}\A^n \nonumber \\ & &
 - 6A_i\A_{\m} + K_{i\m}A_{jk}\A^{jk}
+ R^{j\;k}_{\;n\;i}R^{\ell}_{j\m k}A_{\ell}\A^n \nonumber \\ & &
 + 2R^{\ell}_{j\m i}A_{\ell n}\A^{jn}
+ R^{\ell}_{j\m k}A_{i\ell}\A^{jk} + (D_iR^{\ell}_{j\m k})A_{\ell}\A^{jk}
\nonumber \\ & &
 + 2R^k_{n\m i}A_k\A^n ]\D_\mu z^i\D^\mu\z^{\m}
- 6M^2_\psi K_{i\m}\D_\mu z^i\D^\mu\z^{\m} \big\}.
\eea
Finally,
\begin{equation}
4\L_i\A^iAe^{-K} +{\rm h.c.}= 4\(- e^{-2K}A_{ij}\A^i\A^jA  +
e^{-K}\D_\mu z^i\D^\mu z^j A_{ij}\A +{\rm h.c.}\)$$
$$ + 16A_i\A^iA\A + 8e^{-K}\D^\mu z^i\D_\mu\z^{\m}(\A_iA_{\m} + K_{i\m}A\A).
\end{equation}
Then evaluating $\L_1 - \L_r + \Delta_r\L_1 - \Delta_K\L_1 -
4\(\L_i\A^iAe^{-K} +{\rm h.c.}\)$ yields the result given in (3.6).

\subsection{Errata}
\setcounter{equation}{0}

In this appendix we list errata for references~\cite{josh} and~\cite{noncan}.
In both of these papers the term:
$$\L_q \ni - {x\over 2}h^\rho_\rho F^{\mu\nu}\D_\mu\hA_\nu + x
h^\mu_\nu F_{\mu\rho}\(\D^\nu\hA^\rho - \D^\rho\hA^\nu\) $$
was inadvertently omitted from the quantum Lagrangian, and graviton-Yang-Mills
ghost mixing was neglected; this will be corrected in~\cite{future}.

\subsubsection{Corrections to Ref. 4}

\def\tilG{{\tilde G}}
\def\notG{\not{\hspace{-.065in}\tilG}}
\def\tG{{\tilde G}_{\nu\mu}}
\def\tD{{\tilde D}}
\def\ta{{\tilde A}}
\def\tm{{\tilde M}}
\def\tcF{{\tilde{\cal F}}}
\def\N{\cal N}

\def\thefootnote{\arabic{footnote}}

\def\theenumi{\alph{enumi}}
\begin{enumerate}
\item The D-term is missing from the tree level bosonic Lagrangian
\cite{cremmer} in Eq.(1.8):
\begin{equation}
{1\over \sqrt{g}}\L_B \ni -{1\over 2}{\rm Re}f^{-1}_{ab}\G_i(T^a)^i_jz^j
\G_k(T^b)^k_{\ell}z^{\ell}.\end{equation}

\item A factor $e^{\G/2}$ is missing from the last term in the first line of
(1.11).  The signs of the last term in the fourth line and the second term of
the last line of the same equation should be changed.

\item Eqs.(2.33) and (2.34) should read, respectively:
\bea
 d_\mu &\to& ip_\mu + \tG{\partial\over \partial p_\nu} \nonumber \\
& \equiv& ip_\mu - \sum^{\infty}_{n=1}{n\over (n+1)!}(d_{\mu_1}\cdots
d_{\mu_{n-1}}G_{\mu_n\mu}){(-i)^n\partial^n\over \partial p_{\mu_1}\cdots
\partial p_{\mu_n}},\eea
\begin{equation}  F\to \hF = \sum^{\infty}_{n=0}{1\over n!}(d_{\mu_1}\cdots
d_{\mu_n}F){(-i)^n\partial^n\over \partial p_{\mu_1}\cdots
\partial p_{\mu_n}}.\end{equation}

\item Eq.(2.46) should read:
\bea
\L^{\rm aux}_{\rm reg} &=& {1\over 32\pi^2}\Tr\bigg\{2\mu^2{\tilde M}^2\ln2
\nonumber \\
 & &
 + {1\over 2}\[({\tilde M}^4 + {\tilde \D}^2{\tilde M}^2) - {1\over 2}
(e^2S^2 + [K+Q]^2)]\ln(2\mu_0^2/\mu^2)\]\bigg\}.
\nonumber \\ & & \quad \eea

\item The sign of the gauge connection in (2.48), (3.84) and (3.104) is
incorrect.

\item A term $8\tD^\mu z^i\tD_\mu z^{\jbar}V_{i\jbar} + 8\pp^\mu \pp_\mu\s
V_{s\s}$ is missing from (2.67).

\item Eq. (2.79) should read
\begin{equation}
K^2 = 4\(\tD_\mu\tF^{\mu\rho}\)^a\(\tD^\nu\tF_{\nu\rho}\)_a +
2\(\tD_\sigma\tF_{\mu\nu}\)^a\(\tD^\sigma\tF^{\mu\nu}\)_a.
\end{equation}

\item The sign of part of the gaugino connection is incorrect. Eq.(3.84)
should read
\begin{equation}  (d_\mu)^b_c = \delta^b_c(\partial_\mu + i\Gamma_\mu\gamma_5)
+\epsilon^b_{ca}\ta^a_\mu - {i\over 2}(1/{\rm Re}f)^{ba}{\tilde \D}_\mu
({\rm Im}f_{ac})\gamma_5,\end{equation}
and (3.89) should read
\begin{equation}  (L_\mu)^b_c = - {1\over 2}(1/{\rm Re}f)^{ba}{\tilde \D}_\mu
({\rm Im}f_{ac})\end{equation}
As a consequence of this sign error, there are errors in the $\partial_\mu s$
terms in the final equations (4.1) and (4.2) of I.  The correct result will be
given in~\cite{future}.

\item The right hand sides of (3.106), (3.107) and (3.112) should
be multiplied by ${1\over2}$.

\item
The sign of the right hand side of the last line of (3.91) is
incorrect, and (3.120) should should read:
\begin{equation}
Z\tM = \pmatrix{0&0&0\cr 0&0& -(\tm^{\lambda\chi})_{\mu\nu}\cr 0&
(\tm^{\lambda\chi})_{\mu\nu}&0\cr}.
\end{equation}
This sign error modifies the relative coefficients of (3.174) and (3.75) for
the
terms containing $\tm^{\chi\lambda}$.

\item A term is missing from (3.133):
\begin{equation}
(F^+_{\mu\nu})^{\ibar}_{\jbar}(F^{+\mu\nu})_{\ibar}^{\jbar} \ni
(f^{\mu\nu}f_{\mu\nu})^{\ibar}_{\ibar}. \end{equation}
Corresponding terms that were omitted from the final equations of I will be
given in \cite{future}.

\item The paragraph after Eq.(3.145) should read:

We also need to expand the remaining terms in (3.19).  Only the second
term, being of order $M^2$, will yield both quadratically and and
logarithmically divergent corrections.  The other terms yield only
logarithmically divergent terms.  Again, after using \ldots

\item The signs of $\hM_\psi$ in (3.141) and of $M_\psi,N_\psi$ in
(3.149--152) should be changed; this does not affect the final result.

\item The subscripts $a$ and $\mu$ on $\notG^{\psi\lambda}$
in Eqs.(3.159) and (3.160) should be interchanged.

\item The right hand side of (3.165) should read:
\begin{equation}
\ln(\gamma_\mu\Delta^{\mu\nu}_\theta\gamma_\nu) - \ln(-2/\notp),
\end{equation}
and the second line of the same equation should read:
\begin{equation} -{1\over 2}M_\psi\{\cdots \end{equation}

\item In Eqs.(3.174)--(3.176), $G^{\chi\lambda}$ should be replaced
everywhere by $G^{\psi\lambda}$.  In addition, the subscripts $a$ and
$\sigma$ on $(G^{\psi\lambda}_{\mu\nu})$ in (3.174) should be interchanged,
the denominator of the second term on the right hand side of
(3.176) should be $48p^4$, and the left hand side of (3.176) should be
multiplied by ${1\over 4}$.

\item The last line of (3.192), and the last term in the fourth line of
that equation, should be removed.
\end{enumerate}
\subsubsection{Corrections to Ref. 5}

\begin{enumerate}
\item Eqs. (30) and (31) should read, respectively
\begin{equation}
y^{\alpha\beta}_i = 2k\kappa\(D_{{\tilde\phi}^i}\ln{\tilde e}\)
\[P^{\alpha\beta,\delta\sigma}\eta^{\gamma\epsilon} + {1\over 4\kappa}
\eta^{\gamma\alpha}\eta^{\epsilon\beta}\eta^{\delta\sigma}\]
\Tr\tcF_{\epsilon\sigma}\tcF_{\gamma\delta},$$
$$ K^{\alpha\beta,\epsilon}_a = -4\kappa
\[P^{\alpha\beta,\delta\sigma}\eta^{\gamma\epsilon} + {1\over 4\kappa}
\eta^{\gamma\alpha}\eta^{\epsilon\beta}\eta^{\delta\sigma}\]
\[\(\tD_\sigma\tcF_{\delta\gamma}\)_a - [{\tilde \nabla},\ln{\tilde e}]
\(\tcF_{\delta\gamma}\)_a\],\end{equation}
\item The sentence before (44) should read:

Also, $-2\hf^i\eta^\nu_\alpha(C_\mu)^\alpha_{bi}(\hat{\D}^\mu)^b_a
{\hat{\cal{A}}}^a_\mu=\ldots$

\item The signs of the $C^\alpha C_\alpha$ terms in (46) should be changed.
\item There is a $\pp^\mu\ln{\tilde e}\Delta_{\mu\nu}
\pp^\nu\ln{\tilde e}$ term missing from (53).
\item The last sentence of the paragraph following (53) should read

\ldots where $N'$ is $N$ without the $\Omega$ and $C$ terms \ldots

\item The following corrections to (58) should be made:
\begin{itemize}
\item Replace ${3\over 4}N^{'ab}_{\mu\nu}\delta_{ab}\eta^{\mu\nu}$ by
$\(N-\Omega -{1\over 4}N'\)^{ab}_{\mu\nu}\delta_{ab}\eta^{\mu\nu}$.
\item Replace $-{3\over 2}\S^a_{\mu i}\S^b_{\nu j}Z^{ij}\delta_{ab}
\eta^{\mu\nu}$ by $-\(2\S^a_{\mu i}\S^b_{\nu j}-
{1\over 2}\S^{'a}_{\mu i}\S^{'b}_{\nu j}\)Z^{ij}\delta_{ab}\eta^{\mu\nu}$.
\item Replace ${1\over 2}(N'_{\mu\nu}N^{'\nu\mu})^{ab}\delta_{ab}$ by
$\([N-\Omega]_{\mu\nu}[N-\Omega]^{\nu\mu}
-{1\over 2}N'_{\mu\nu}N^{'\nu\mu}\)^{ab}\delta_{ab}.$
\item Replace $2N'_{\mu\nu}\Omega^{\mu\nu}$ by
$2(N-\Omega)_{\mu\nu}\Omega^{\mu\nu}$.
\end{itemize}

\item The text following (58) should read ``\ldots where $\S' = S+s$, and
$N'$ is given by $N$ in (46)
without the $\Omega$ and $C^2$ terms, \ldots''.
\item The second sentence of the Appendix should read ``The space-time metric
$g_{\mu\nu}$ has the flat limit $\eta_{\mu\nu}=$ diag$(1,-1,-1,-1)$.''
\item Replace $e$ by ${\tilde e}$ in the definition of $Q$, eq. (A1).

\end{enumerate}

\end{document}